# Transfer learning relaxation, electronic structure and continuum model for twisted bilayer MoTe$_2$

Check for updates

Ning Mao[1,7], Cheng Xu [2,3,7], Jiangxu Li[2], Ting Bao [3], Peitao Liu [4], Yong Xu [3], Claudia Felser[1], Liang Fu [5] & Yang Zhang [2,6] ✉

Large-scale moiré systems are extraordinarily sensitive, with even minute atomic shifts leading to significant changes in electronic structures. Here, we investigate the lattice relaxation effect on moiré band structures in twisted bilayer MoTe$_2$ with two approaches: (a) large-scale plane-wave basis first principle calculation down to 2.88°, (b) transfer learning structure relaxation + local-basis first principles calculation down to 1.1°. We use two types of van der Waals corrections: the D2 method of Grimme and the density-dependent energy correction, and find that the density-dependent energy correction yields a continuous evolution of bandwidth with twist angles. Based on the above results. we develop a complete continuum model with a single set of parameters for a wide range of twist angles, and perform many-body simulations at $v = -1, -2/3, -1/3$.

Recent experiments on twisted bilayer MoTe$_2$ (tMoTe$_2$) reported the observation of the fractional quantum anomalous Hall (FQAH) effect at fractional fillings $\nu = -\frac{2}{3}$ and $-\frac{2}{5}$ of the moiré band[1–4]. The realization of the FQAH effect in twisted transition metal dichalcogenide bilayers (tTMDs) was theoretically proposed[5–7] as a consequence of band topology[8] and electron interaction. Specially, spontaneous ferromagnetism and electron correlation in spin-valley polarized Chern band lead to the emergence of fractional Chern insulators (FCI) that exhibit the FQAH effect at zero magnetic fields. The observed FQAH effect in tMoTe$_2$ is remarkably robust, existing over an unexpectedly wide range of twist angles and persisting up to 2 K[3]. The experimental realization of the long-sought fractional quantum Hall effect at zero magnetic field[9–18] not only expands the realm of fractionalized topological phases, but also holds promise for anyon-based topological quantum computations[19,20].

While theoretical studies have provided important insights into the FQAH effect in tTMDs, the underlying moiré band structures of tMoTe$_2$ over the experimentally accessible twist angle range has not been systematically studied. A number of first-principles studies report different bandwidths at the commensurate angle $\theta = 3.89°$, ranging from 9 to 18 meV for the lowest moiré band[21–23]. Importantly, lattice relaxation at the moiré length scale can significantly impact the band structure and even the band topology. While the effect of the out-of-plane corrugation has been considered, the in-plane lattice relaxation and the effect of the resulting strain field have not been incorporated into the continuum model. The strain-induced pseudomagnetic field, as well as higher-harmonic moiré potentials, strongly affect higher moiré bands, and therefore are crucial for studying band-mixing FCI states[23–25] and interaction-induced phases at higher filling factors. Finally, first-principles electronic structure calculations for twist angles below $\theta = 3.89°$ are entirely lacking. The accuracy of the continuum model at small twist angles remains to be assessed.

In this work, we perform extensive first-principles simulations to study moiré lattice relaxation and electronic structures. Our calculations encompass a wide range of twist angles, reaching as small as 2.88° using plane-wave basis and 1.1° using transfer learning technique and local basis. In addition to interlayer corrugation, we observe significant in-plane displacement[26–30] reaching around 0.5 Å at small angles. To capture the significant effect of lattice relaxation, we extend the continuum model to include second harmonic moiré potentials and pseudo-magnetic field up to 250 T from in-plane strain[27]. Remarkably, all four topmost moiré valence bands over the entire range of experimentally relevant twist angles ($\theta = 2.6° - 5°$[1–4]) are accurately reproduced by our continuum model, with a single set of parameters. With the transfer-learning model, we further calculate the topological edges states and Wilson loop around 2°, revealing a series of $C = 1$ Chern bands. These findings serve as the foundation point for even-denominator non-Abelian states.

## Results

### Two types of van der Waals corrections

For accurate lattice relaxations in two-dimensional (2D) multi-layer systems, it is essential to incorporate van der Waals (vdWs) dispersion corrections into the total energy, potential, interatomic forces, and stress tensor calculations. The choice of vdWs corrections, therefore, influences the lattice parameters of unit cells and the interlayer distances. Typically, vdWs corrections fall into two categories: (a) charge-density independent methods such as DFT-D2/D3 and (b) charge-density-dependent methods. The latter category accounts for charge-density variations in vdWs contributions of atoms influenced by their local chemical environments.

The DFT-D2 method[31] adds an empirical single-shot dispersion correction to the conventional density functional theory (DFT) calculations. The correction term for the dispersion energy $E_{disp}$ is given by

$$E_{disp} = -s_6 \sum_{i=1}^{N-1} \sum_{j=i+1}^{N} \frac{C_{6,ij}}{R_{ij}^6} \cdot f_{damp}(R_{ij}). \quad (1)$$

Here, $s_6$ denotes a global scaling factor that only depends on the density functional used, $N$ is the number of atoms in the system, $C_{6,ij}$ are the dispersion coefficients for the atom pair $(i,j)$, and $R_{ij}$ is the distance between atoms $i$ and $j$. The damping function $f_{damp}$ is used to avoid the divergence of the dispersion term at short interatomic distances. $C_{6,ij}$ and $f_{damp}$ are determined by the local geometry, which is unrelated to the self-consistent iteration.

Unlike the D2 method, the density-dependent screened Coulomb (dDsC) method[32,33] involves a density-dependent screening function to modulate the Coulomb interaction, which allows for a more realistic representation of vdWs interactions as a function of the local chemical environment. The correction term for dDsC can be expressed by

$$E_{disp} = -\sum_{i=1}^{N-1} \sum_{j=i+1}^{N} \frac{C_{6,ij}}{R_{ij}^6} \cdot f_{damp}(bR_{ij}). \quad (2)$$

The key difference between dDsC and DFT-D2 lies in the damping function $f_{damp}$, which is associated to the key component $b$ (damping factor) for dDsC. This damping factor can be determined by the local electron density, the gradient of the electron density, and other environment-specific parameters. Therefore, it is particularly useful for systems (e.g., strongly correlated moiré systems studied here) where vdWs interactions are sensitive to the local electronic environment.

For untwisted bulk structures, these two vdWs correction methods often give similar results. As shown in Supplementary Note 1, for the bulk-$MoTe_2$, the lattice constants and the vertical layer distances predicted by both DFT-D2 and dDsC methods agree well with the experimental results ($a$ = 3.52 Å and $d$ = 6.99 Å)[34,35]. However, for the moiré superlattice system, the dDsC method yields more reliable structure relaxation, since the rich local chemical environments such as position-dependent electrical dipoles appearing in the moiré superlattice are better described by dDsC.

### Large-scale DFT and lattice relaxation effect

Making use of the initial moiré structure generated by deep potential molecular dynamics (DeePMD)[36], large-scale structural relaxations can be achieved at a significantly reduced computational cost. Remarkably, the relaxation of $\theta$ = 2.88° twisted structures comprising 2382 atoms was completed in just 5 h with 17 DFT ionic steps using DeePMD-generated structure in four NVIDIA H100 GPUs. The self-consistent calculation and band diagonalization of this 2382-atom system (IBAND = 17,160 and plane-wave number 11,469,590) can be done within 80 min in 20 NVIDIA H100 GPUs, showcasing the massive speedup of the GPU platform for large-scale first principle simulation.

To demonstrate the relaxation effect in the $tMoTe_2$, we compare the relaxed moiré structures with twist angles 3.89° and 2.88°. First, there is a big variation in the interlayer spacing (ILS) (Fig. 1), indicating a large structural transformation. For $tMoTe_2$ with a twist angle of 3.89° (Fig. 1a), the maximum ILS observed is 7.8 Å. This occurs in the MM region, where the Te/Mo atoms of the top layer are aligned directly above those in the bottom layer, resulting in an energy increase in this area due to the strong repulsion. The minimum ILS is 7.0 Å, which is observed in the MX region where the Mo atoms of the top layer stack over the Te atoms of the bottom layer. Figure 1b shows the ILS for 2.88°$tMoTe_2$ exhibiting a clear domain wall connecting MM regions, which becomes more significant at lower twist angles, as shown in Supplementary Figs. 2–5.

Concerning the intralayer strain, both structures exhibit similar behaviors. As depicted in Fig. 1c, d, the in-plane displacement pattern displays a helical chirality, with the amplitude intensifying as the twist angle diminishes. We observe a large displacement up to 0.5 Å for $\theta$ = 2.88°, which generates a pseudomagnetic field up to 200 $T$ (see Supplementary Note 3).

### Symmetry analysis of moiré band structures

The space group of the relaxed structures is $P321$ (No. 150), whose point group is generated by a twofold rotational symmetry along $y$ axis ($C_{2y}$), and three-fold rotational symmetry along $z$ axis ($C_{3z}$). In the crystal momentum space, the $C_{2y}$ symmetry only protects twofold degeneracies at the invariant lines or points within the Brillouin Zone, as defined by the relation $C_{2y}\mathbf{k} \to \mathbf{k}$. Within this invariant space, the Hamiltonian commutes with the symmetry operation, allowing it to be block-diagonalized into two distinct sectors, each characterized by unique eigenvalues $\pm \pi$. Due to the constraints imposed by the symmetry, a band represented by $e^{i\pi}$ is inherently degenerate with another band represented by $e^{-i\pi}$, forming a doubly degenerate band structure. Consequently, the only lines that encapsulate the $C_{2y}$ symmetries within the two-dimensional Brillouin zone are the $\Gamma K$ lines (satisfying $2\mathbf{k}_1 + \mathbf{k}_2 = 0$). When considering the $C_{3z}$ rotational symmetry, the lines that meet the conditions $k_1 + 2k_2 = 0$ and $k_1 - k_2 = 0$ also emerge as symmetry-invariant lines. As a result, bands along the $\Gamma K$ and $MK$ lines are always doubly degenerate, while a clear splitting is observed along the $\Gamma M$ line, as shown in Fig. 2.

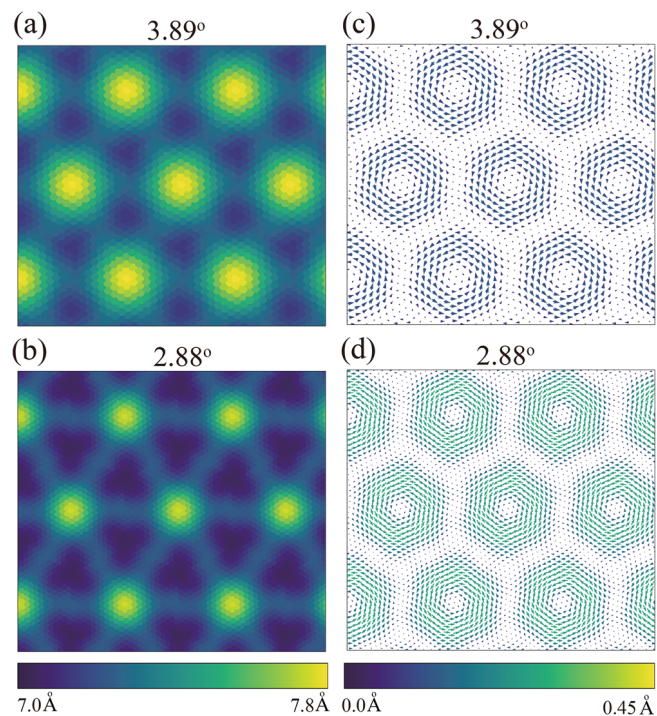

**Fig. 1 | Lattice relaxation of interlayer and intralayer distances.** The relaxed structures of 3.89° and 2.88° $tMoTe_2$ from density functional theory under density-dependent screened Coulomb van der Waals correction. **a**, **b** are interlayer distances, determined by the distance between Mo atoms at the top and bottom layers. **c**, **d** are intralayer displacements, indicating the in-plane displacements after relaxations for Mo atoms at the top layer.

In Fig. 3, we plot the angle-dependent bandwidth and direct gap using two types of vdW corrections. At twist angle $\theta = 3.89°$, D2 type of vdW correction gives rise to a narrow bandwidth as 12 meV, which is close to previous calculation using local-basis SIESTA package[37] and D2 correction[21]. While under the dDsC type of vdW correction, we obtain the bandwidth as 18 meV, and the overall trend of angle-dependent bandwidth follows a parabolic continuum behavior with a single set of parameter, as we will discuss later. At the smallest calculated twist angle $\theta = 2.88°$, the width of the top moiré valence band reduced to 6 meV.

## Complete continuum model

We now introduce a more comprehensive continuum model to depict the moiré band structure. The key low-energy states originate from the hole bands in the $K$ and $K'$ valleys of the two MoTe$_2$ layers. Considering that these valleys are connected through time-reversal symmetry ($\mathcal{T}$), analyzing one valley is sufficient to infer the band structure. For tTMD systems with rotational ($C_{3z}$) and layer-exchange symmetry ($C_{2y}\mathcal{T}$), we derive the following form:

$$\hat{H} = \begin{bmatrix} -\frac{(\hat{k}-K_t+eA)^2}{2m^*} + \Delta_t(r) & \Delta_T(r) \\ \Delta_T^\dagger(r) & -\frac{(\hat{k}-K_b-eA)^2}{2m^*} + \Delta_b(r) \end{bmatrix} \quad (3)$$

with:

$$\Delta_t(r) = 2V_1 \sum_{i=1,3,5} \cos(g_i^1 \cdot r + l\phi_1) + 2V_2 \sum_{i=1,3,5} \cos(g_i^2 \cdot r)$$

$$\Delta_T = w_1 \sum_{i=1,2,3} e^{-iq_i^1 \cdot r} + w_2 \sum_{i=1,2,3} e^{-iq_i^2 \cdot r} \quad (4)$$

$$A(r) = A(a_2 \sin(G_1 \cdot r) - a_1 \sin(G_3 \cdot r) - a_3 \sin(G_5 \cdot r))$$

where $\hat{k}$ is the momentum measured from the $\Gamma$ point of single layer MoTe$_2$, $K_t(K_b)$ is high symmetry momentum $K$ of the top (bottom) layer, $\Delta_t(r)(\Delta_b(r))$ is the layer dependent moiré potential, $\Delta_T(r)$ is the interlayer tunneling, $G_i$'s are moiré reciprocal vectors, $A(r)$ is the strain-induced gauge field which gives a periodic pseudomagnetic field[27,38].

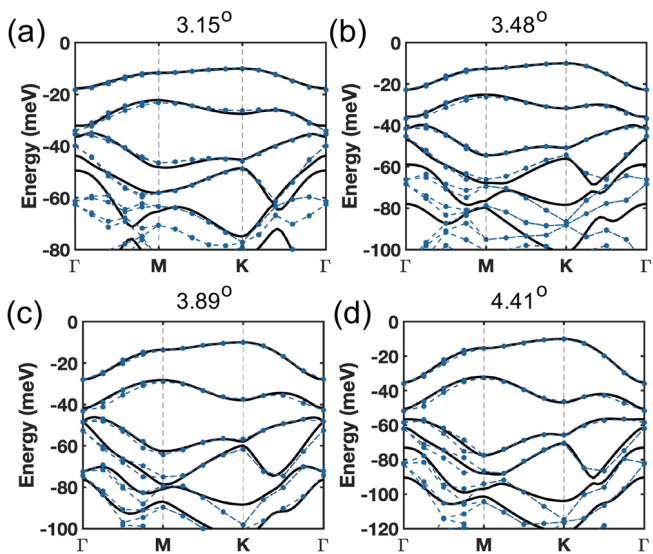

**Fig. 2 | Fitting results from the continuum model.** The comparative analysis of the band structures for twist angle **a** 3.15°, **b** 3.48°, **c** 3.89°, and **d** 4.41°, respectively. Blue points/lines illustrate the results from density functional theory calculations, while the black line represents the fitting results from the continuum model.

$B(r) = B \sum_{i=1,3,5} \cos(G_i \cdot r)$. $g_i^1$ and $g_i^2$ represent the momentum differences between the nearest and second-nearest plane-wave bases within the same layer. Similarly, $q_i^1$ and $q_i^2$ denote the momentum differences between the nearest and second-nearest plane-wave bases across different layers. $a_1, a_2, a_3$ are the moiré lattice vectors. The relations between different wave vector can be given by $G_1 = \frac{4\pi}{\sqrt{3}a_M}(\frac{1}{2}, -\frac{\sqrt{3}}{2})^T$, $G_3 = \frac{4\pi}{\sqrt{3}a_M}(\frac{1}{2}, \frac{\sqrt{3}}{2})^T$, $G_5 = \frac{4\pi}{\sqrt{3}a_M}(-1, 0)^T$, $q_1^1 = \frac{4\pi}{3a}2\sin(\frac{\theta}{2})(0,1)^T$, $q_2^1 = C_3 q_1^1$, $q_3^1 = C_3^2 q_1^1$, $q_1^2 = q_1^1 - G_5$, $q_2^2 = C_3 q_1^2$, and $q_3^2 = C_3^2 q_1^2$.

To obtain accurate parameters in the continuum model, we perform large-scale calculations with dDsC vdWs corrections (IVDW = 4), then fit the DFT moiré band structure at 3.15° to obtain the following continuum parameter: $m^* = 0.62 m_e$, $V_1 = 10.3$ meV, $V_2 = 2.9$ meV, $w_1 = -7.8$ meV, $w_2 = 6.9$ meV, $\phi_1 = -75°$, $\Phi/\Phi_0 = 0.737$. ($\Phi_0$ is the quantum flux, $\Phi$ represents the value of flux in the moire unit cell). The continuum model parameters with D2 vdW corrections (IVDW = 10) are presented in the Supplementary Note 4. In our subsequent analysis of the continuum model, we will utilize the parameter from the IVDW = 4, as it provides the more reliable structure relaxation previously discussed. Employing these parameters, we are now equipped to solve the moiré band structures at various small twist angles.

Next, we examine the topology of these moiré bands from 1.6° to 5°. At twist angles below 2.5°, the Chern numbers for the top three bands, as calculated using the continuum model, are 1, −1, 0, as shown in Fig. 4c (see Supplementary Note 2). For greater twist angles, these Chern numbers change to 1, 1, −2. We emphasize that the arrangement of Chern numbers for $\theta > 2.83°$ is in agreement with experimental data. So far, in all experiments where twist angle $\theta$ ranges between 3.5°–3.9°, both[1,2] the Hall conductance and the reflective magnetic circular dichroism increase once the doping exceeds $\nu = -1$. And a double quantum spin Hall effect has been observed at $\nu = -4$. These results suggest that the second band shares the same Chern number as the first band.

We additionally verify the trace condition for the uppermost moiré band. The band's geometry is encapsulated in the quantum geometry tensor:

$$\eta^{uv} := A_{BZ} \langle \partial^u u_k | (1 - |u_k\rangle\langle u_k|) | \partial^v u_k \rangle \quad (5)$$

where $A_{BZ}$ is the area of the Brillouin zone. The symmetric and antisymmetric parts of the quantum geometry tensor give the Berry curvature ($\Omega(k) = -2\text{Im}(\eta^{xy})$) and quantum metric ($g^{uv}(k) = \text{Re}(\eta^{uv})$). To quantify the geometric properties, one can calculate two figures of merits[39–41]:

$$\sigma_F := \left[\frac{1}{A_{BZ}} \int d^2 k (\frac{\Omega(k)}{2\pi} - 1)^2\right]^{\frac{1}{2}}$$

$$T := \frac{1}{A_{BZ}} \int d^2 k \left[tr(g(k)) - \Omega(k)\right], \quad (6)$$

where $\sigma_F$ describes the fluctuations of Berry curvature and the $T$ quantifies the violation of the trace condition. When both $\sigma_F$ and $T$ tend towards 0, it becomes possible to exactly map the Chern band to a Landau-level problem, allowing for an intuitive understanding of the fractional state. We calculate the values of these parameters in relation to the twist angle, as depicted in Fig. 4d.

## Transfer learning structure relaxation

In order to resolve the problem of structural relaxation, we adopt the ab initio DeePMD method, which combines the first-principles accuracy and empirical-potential efficiency for large-scale systems[36]. We begin with $3 \times 3 \times 1$ MM, MX, and XM configurations, along with 28 distinct intermediate transition states, all of which have been relaxed with a fixed volume. For each one of the 31 configurations, we introduced random perturbations to generate 200 distinct structures. The random perturbations are applied to both the atomic coordinates, drawing values from a uniform distribution spanning [−0.01 Å, 0.01 Å], and the lattice constants, guided by a deformation matrix that is constructed from a distorted identity matrix spanning [−0.03, 0.03]. Besides, we conduct the 20 fs ab initio molecular dynamics to gather VASP-calculated energy, force, and virial tensor, which constitute the entirety of the initial training set.

Next, we train the initial neural network model through the initial training set, and run molecular dynamics simulations for different pressures (−100 to 10000 bar) and temperatures (10 to 500 K). A bunch of trajectories are generated in this process, and we label them as the failure, candidate, or accurate configurations according to the model deviation: $\sigma_f^{max} = \max \sqrt{\langle |F_i - \langle F_i \rangle|^2 \rangle}$. During the process, 3 to 200 candidate configurations will be selected to perform the self-consistent DFT calculations, and the data will be collected for the training process of next-iteration.

Although the neural network model shows effective performance in the IVDW-10 correction, it does not yield successful results in the IVDW-4 correction, largely attributed to the complex dependencies on charge density. To address the issue, we augment our training datasets with comprehensive data from twisted structures of 2.88°, encompassing 118 sets of forces, energies, and virial information, as illustrated in Fig. 5. Leveraging the principles of transfer learning, we strategically froze the parameters within the embedding layers while focusing on training the hidden and output layer. This approach significantly improves the performance of the pre-trained model, enabling it to adapt more effectively to the complexities of IVDW-4.

## Conclusion

In this paper, we delve deeply into the lattice relaxation and single-particle of the $t$MoTe$_2$ system. We present a comprehensive exploration of the moiré band structure under two types of vdW corrections, where we harness the power of large-scale DFT calculations together with transfer learning and GPU acceleration. Built on angle-dependent moiré band structures, we construct a more complete continuum model including higher-harmonic potential and strain-induced gauge field. Our calculations reveal that, at experimentally pertinent twist angles, the intralayer displacement induces a sizeable gauge field, and top two moiré bands consistently display nontrivial Chern numbers.

The continuum model parameters have a strong impact on interaction-induced phases in $t$MoTe$_2$, as shown by previous numerical studies[21–25,42–46]. With the continuum model fitted from D2 type of vdW correction[21], integer quantum anomalous Hall effect only appears at large dielectric constant $\epsilon > 15$[24,25], and $\nu = -\frac{2}{3}, -\frac{1}{3}$ are both found to be FCIs[21,43]. With the continuum model fitted from dDsC type of vdW correction[22,23], the integer quantum anomalous Hall effect has been shown to occur at experimentally studied twist angles[24,25], while $\nu = -\frac{2}{3}$ and $\nu = -\frac{1}{3}$ are FCI and charge-density wave states, respectively[22,25].

Note: Upon the completion of this work, a related work appeared[47], which overlaps with some of our calculations with IVDW = 10.

## Method
### Plane-wave basis first principle calculations
The large-scale plane-wave basis first principle calculations are carried out with Perdew–Burke–Ernzerhof (PBE) functionals using the Vienna Ab initio simulation package (VASP)[48–50]. We chose the projector augmented wave potentials, incorporating six electrons for each of the Mo and Te atoms. During the structural relaxation, we set the plane-wave cutoff energy and the energy convergence criterion to 250 eV and $1 \times 10^{-6}$ eV, respectively. Larger energy cutoff of 350 and 500 eV has been tested for $\theta = 3.89°$, which leads to less than 1 meV change in the bandwidth of topmost valence band. The structure is fully relaxed when the convergence threshold for the maximum force experienced by each atom is less than 10 meV/Å.

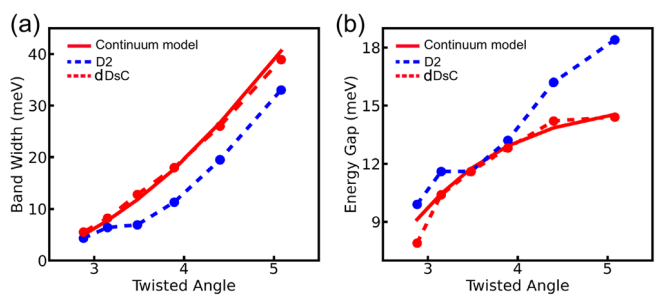

**Fig. 3 | Variation of bandwidth and direct band gap under different van der Waals corrections. a** Variation in the bandwidth of the first and second moiré bands versus diverse twist angles. **b** Variation in direct band gap between the first two bands along the high symmetry lines in Fig. 2. The blue and red dashed curves denote the D2 (IVDW-10) and density-dependent screened Coulomb (IVDW-4) van der Waals corrections, while the red solid line represents the results from continuum model.

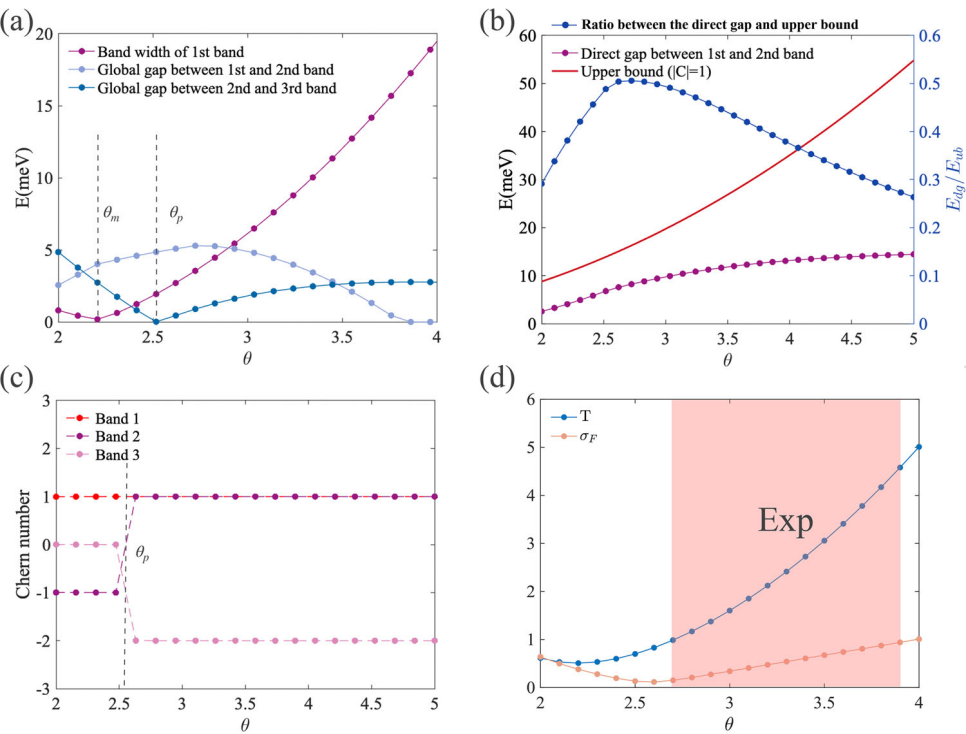

**Fig. 4 | Band topology of the continuum model. a** The bandwidth and the global gap from the continuum model. Gap closure happens between the second and third band around $\theta_p \approx 2.5°$. The bandwidth of the first band is denoted with the purple curve, which attains zero at the magic angle $\theta_m \approx 2.2°$. **b** The direct gap between the first and second band and we note it always below the upper bound[54]. The red and purple curves denote the direct band gap ($E_{dg}$) and the upper bound ($E_{ub}$), respectively. The blue curve represents the ratio between $E_{dg}$ and $E_{ub}$. **c** The Chern number of the top three bands. There is a topological phase transition around 2.5° owing to gap closure between the second and third bands. **d** $T$ describes the violation of trace condition, and $\sigma_F$ reflects the fluctuation of Berry curvature. We show them as a function of twist angles. The red region denotes the range of angles where fractional quantum anomalous Hall effect are observed in transport experiments.

**Fig. 5 | Scheme of transfer learning.** The initial dataset consists only of non-twisted structures, and we use density functional theory calculations to collect the energy and force data. A neural network is then trained with DeePMD. However, this neural network struggles to accurately predict the forces in twisted structures. To address this, we build another dataset with twisted structures and use the energy and force data from this new dataset to perform a transfer learning based on the original neural network. This new neural network demonstrates a significantly improved ability to predict forces accurately.

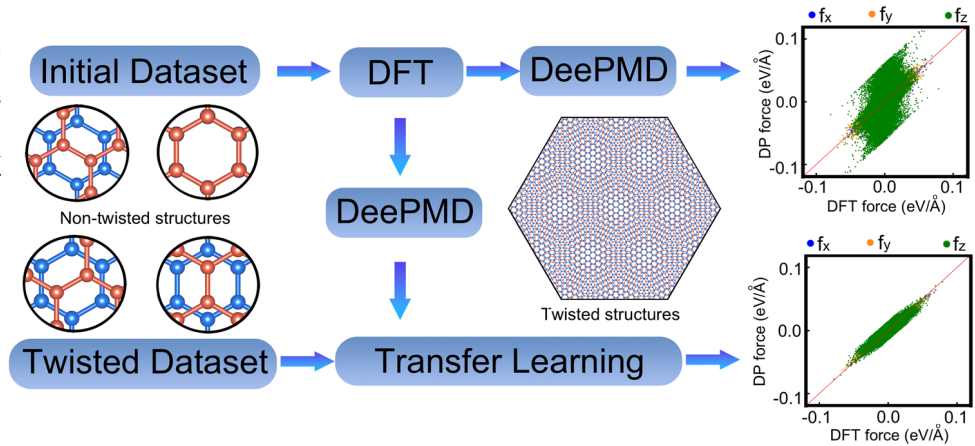

### Local-basis first-principles calculations
Apart from the calculation utilizing the plane-wave basis, our DFT study on $t$MoTe$_2$ has also been performed under the pseudo atomic orbital (PAO) basis. Using the relaxed atomic structures from VASP and DeePMD, we use OpenMX package[51,52] with PAOs chosen to be $Mo7.0$-$s3p2d1$ (7.0 means the cutoff radius is 7.0 Bohr, $s3p2d1$ means three sets of $s$-orbitals, two sets of $p$-orbitals and 1 set of $d$-orbitals, summed up as $3 \times 1 + 2 \times 3 + 1 \times 5 = 14$ atomic orbitals for each Mo atom) and $Te7.0$-$s3p2d2$ to conduct the self-consistent calculation and obtain the band structure. The PBE exchange-correlation functional and the norm-conserving pseudopotential[53] are employed in the calculation with single $\Gamma$ $k$-sampling and convergence criterion no lower than $6 \times 10^{-5}$ Hartree.

### Machine learning workflow
We are using the DeePMD-kit code to train the neural network[36]. Here, we adopt the two-body embedding smooth edition of the DeepPot-SE descriptor, which is constructed by both angular and radial atomic configurations. The cut-off and smooth radius for neighbor searching are set as 8.0 and 2.0 Å, including a maximum number of 100 Mo and 100 Te atoms. Then, we construct a neural network that maps the descriptors to atomic energy, through three embedding layers and three hidden layers of size (25, 50, 100) and (240, 240, 240), respectively. To measure the quality of the neural network, we construct a loss function by a sum of different root means square errors (RMSE):

$$L(p_\epsilon, p_f, p_\xi) = \frac{p_\epsilon}{N}\Delta E^2 + \frac{p_f}{3N}\sum_i |\Delta F_i|^2 + \frac{p_\xi}{9N} \| \Delta \Xi \|^2, \quad (7)$$

where $\Delta E$, $\Delta F_i$, and $\Delta \Xi$ refer to the RMSE of energy, force, and virial, respectively. During the training process, the prefactor $p_f$ decreases from 1000 to 1, $p_\epsilon$ and $p_\xi$ increase from 0.02 to 1. To improve the efficiency of network training, we adopt an exponentially decaying learning rate to minimize the loss function. After 1,700,000 training steps, the learning rate decreases from $1e^{-3}$ to a small value of $3.6e^{-8}$.

### Data availability
The data that support the findings of this study are available from the corresponding authors upon reasonable request.

### Code availability
The code that supports the findings of this study are available from the corresponding authors upon reasonable request.

## Acknowledgements
We are grateful to Tingxin Li, Taige Wang, Trithep Devakul, Fengcheng Wu, and Allan Macdonald for their helpful discussions. Y.Z. thanks Quansheng Wu and Jianpeng Liu for the cross-check on DFT parameters. L.F. and C.F. are partly supported by the Catalyst Fund of the Canadian Institute for Advanced Research. Y.Z. is supported by the start-up fund and the seed grant from the AI Tennessee Initiative at the University of Tennessee Knoxville. The research by J. L. was primarily supported by the National Science Foundation Materials Research Science and Engineering Center program through the UT Knoxville Center for Advanced Materials and Manufacturing (DMR-2309083). The machine learning simulations and large matrix diagonalizations are performed on H100 nodes provided by the AI Tennessee Initiative.

## Author contributions
Y.Z. initiated this project. N.M. performed the first principle calculations, transfer learning structure relaxation, and continuum model calculations, with the help of C.X., J.X.L., and T.B. P.T.L., Y.X., C.F., and L.F. contributed to data analysis. Y.Z., N.M., C.X., and J.X.L. wrote the manuscript with input from all the authors.

## Competing interests
The authors declare no competing interests.

## Additional information

**Supplementary information** The online version contains supplementary material available at
https://doi.org/10.1038/s42005-024-01754-y.

**Correspondence** and requests for materials should be addressed to Yang Zhang.

**Peer review information** This manuscript has been previously reviewed at another Nature Portfolio journal. The manuscript was considered suitable for publication without further review at *Communications Physics*.

**Reprints and permissions information** is available at
http://www.nature.com/reprints

**Publisher's note** Springer Nature remains neutral with regard to jurisdictional claims in published maps and institutional affiliations.